\documentstyle[epsfig,psfig,12pt,emulateapj]{article}
\lefthead{Naik and Paul}
\righthead{High and low state X-ray properties of LMC~X-4}
\slugcomment{Accepted for Publication in The Astrophysical Journal}
\newcommand\sax{{\it Beppo}-SAX}

\received{2003 March 10}
\begin{document}
\title{Timing and spectral studies of LMC~X-4 in high and low states 
with \sax\ : Detection of pulsations in the soft spectral component}

\author{S. Naik\altaffilmark{1,2} \& B. Paul\altaffilmark{1}}
\altaffiltext{1}
{Tata Institute of Fundamental Research, Homi Bhabha Road, Mumbai 400 005, India\\
sachi@tifr.res.in, bpaul@tifr.res.in}
\altaffiltext{2}
{Department of Physics, University College Cork, Cork, Ireland\\
sachi@ucc.ie}

\begin {abstract}

We report here detailed timing and spectral analysis of two Beppo-SAX
observations of the binary X-ray pulsar LMC~X-4 carried out during the
low and high states of its 30.5 days long super-orbital period.
Timing analysis clearly shows 13.5 s X-ray pulsations in the high state
of the super-orbital period which allows us to measure the mid-eclipse
time during this observation. Combining this with two other mid-eclipse
times derived earlier with the ASCA, we derived a new estimate of the
orbital period derivative. Pulse-phase averaged spectroscopy 
in the high and low states shows that the energy spectrum in the 
0.1 -- 10 keV band comprises of a hard power-law, a soft excess, and
a strong iron emission line.
The continuum flux is found to decrease by a factor of $\sim$ 60 in the
low state while the decrease in the iron line flux is only by a factor
of $\sim$ 12, suggesting a different site for the production of the line
emission. In the low state, we have not found any significant increase in
the absorption column density. The X-ray emission is found to come from
a very large region, comparable to the size of the companion star.
Pulse phase resolved spectroscopy in the high state shows a pulsating
nature of the soft spectral component with some phase offset compared to
the hard X-rays, as is known in some other binary X-ray pulsars.

\end{abstract}


\section{Introduction}

LMC~X-4 is an eclipsing, accretion-powered, binary X-ray pulsar in the
Large Magellanic Cloud orbiting around a 20 M$_\odot$ O7 III-V companion 
with an orbital period of $\sim$1.4 day (Kelley et al. 1983, Ilovaisky 
et al. 1984) and pulse period 
of $\sim$13.5 s (\cite{whit:78,li:78,lang:81}). X-ray eclipses with a
1$^d$.4 recurring period were discovered by Li et al. (1978) and White (1978).
X-ray pulsations with a period of 13.5 s were first detected in LMC~X-4 from
{\it SAS 3} observations in 1976 and 1977 only during the infrequent flaring
events by Kelley et al. (1983), who also derived the orbital solution from pulse
arrival delay measurements. During a high state in 1983, {\it EXOSAT}
observations detected the pulsations both during flaring and non-flaring
episodes (\cite{pietsch:85}). Using subsequent pulse timing measurements with 
{\it GINGA}, {\it ROSAT} (\cite{levine:91,woo:96}) and RXTE (\cite{levi:00}),
an orbital period decay with a time scale of $\sim 10^{6}$ yr$^{-1}$ was 
measured, which is similar to the period decay rate in other high mass 
accreting binary pulsars (Cen X-3, SMC X-1 etc.).

Analogous to the well known X-ray pulsar Her~X-1, LMC~X-4 exhibits a periodic 
long term intensity variation at 30.5 day (\cite{lang:81}). The super-orbital
period has recently been found to be decreasing (Paul \& Kitamoto 2002).
In the case of LMC~X-4, the X-ray flux varies by two orders of magnitude
between the high and low intensity states of the super-orbital period.
Varying obscuration by the accretion disk due to precession provides a good 
explanation for the long term periodic intensity variations. The 13.5 s X-ray 
pulsation is not yet detected during the low state of the super-orbital period 
indicating that most of the radiation observed in low state is probably 
reprocessed emission from the stellar wind of the companion star.
The X-ray spectrum during eclipse of the neutron star, which consists only
of scattered/reflected radiation was found to be nearly identical in shape and
strength throughout the super-orbital period, supporting the disk obscuration
hypothesis (\cite{woo:95}). However, observations with the {\it RXTE}
(\cite{hein:99}) did not detect any significant change in the absorption
column density in different phases of the super-orbital period.
One possibility that was pointed out is that in low state, part of the
X-ray emission from the central source is completely blocked by the disk
and the rest is observed with the Galactic absorption. It should be
noted that, for sensitive measurements of absorption column density,
observations with good soft X-ray spectral capability are required.

The X-ray spectrum of LMC~X-4 consists of, in addition to a hard power-law
component, a soft excess below 1 keV and an iron emission line at 6.4
keV (Woo et al. 1996). Iron K shell emission lines in X-ray pulsars are 
usually narrow, but the equivalent width can sometimes be as large as 
$\sim$1 keV or more. These lines are produced by illumination of neutral or 
partially ionized material in accretion disk, stellar wind of high mass 
companion star, material in the form of circumstellar shell, material in 
the line of sight, or accretion column.
Expectedly, the iron emission line parameters are often found to be
correlated to the continuum hard X-ray flux, absorption column density etc.
Using a large number of RXTE/PCA observations of LMC~X-4 made in different
phases of the super-orbital period, a correlation was found between the
line flux and the continuum flux, along with highly variable nature of the
equivalent width of the line during low state (\cite{np:03}). 

It is known that X-ray pulsars those do not suffer from strong absorption
by material in the line of sight show presence of soft X-ray excess above
the extrapolated hard power-law component (\cite{paul:02}, and references
therein). In some pulsars, the soft excess is also found to pulsate, often
with a phase difference with respect to the hard component. Origin of the
pulsating soft component is not yet understood clearly, at least for the
bright pulsars in LMC and SMC. In LMC~X-4, a small Galactic absorption
column density and lack of local absorbing material allows us to probe
the soft X-ray emission with some detail. Observations of LMC~X-4 with ASCA
detected the soft excess but were found to be inconclusive regarding the nature
of the soft component (\cite{paul:02}). The latter is due to a small pulse
fraction of this pulsar and also to the fact that the ASCA spectrometers
are not sensitive at energies where the soft excess dominates the power-law
component in LMC~X-4.

We have carried out detailed timing and spectral analysis of two
observations of LMC~X-4 with the narrow field instruments of Beppo-SAX
during one high and one low intensity
states of LMC~X-4. We aimed to detect pulsations in the Beppo-SAX
observation carried out in the high state and to determine the mid-eclipse
time during this observation. From pulse phase averaged spectral studies
in the two states, we investigate relations of the iron line parameters
with the continuum flux. To investigate the nature of the soft excess,
pulse-phase-resolved spectral analysis has been carried out for the high
state observation. In the subsequent sections we give details of the
observations, the results obtained from the timing and spectral analysis,
followed by a discussion on the results obtained from these two \sax\
observations.

\section{Observations}

\begin{figure*}
\centering
\vskip 9.0cm
\includegraphics{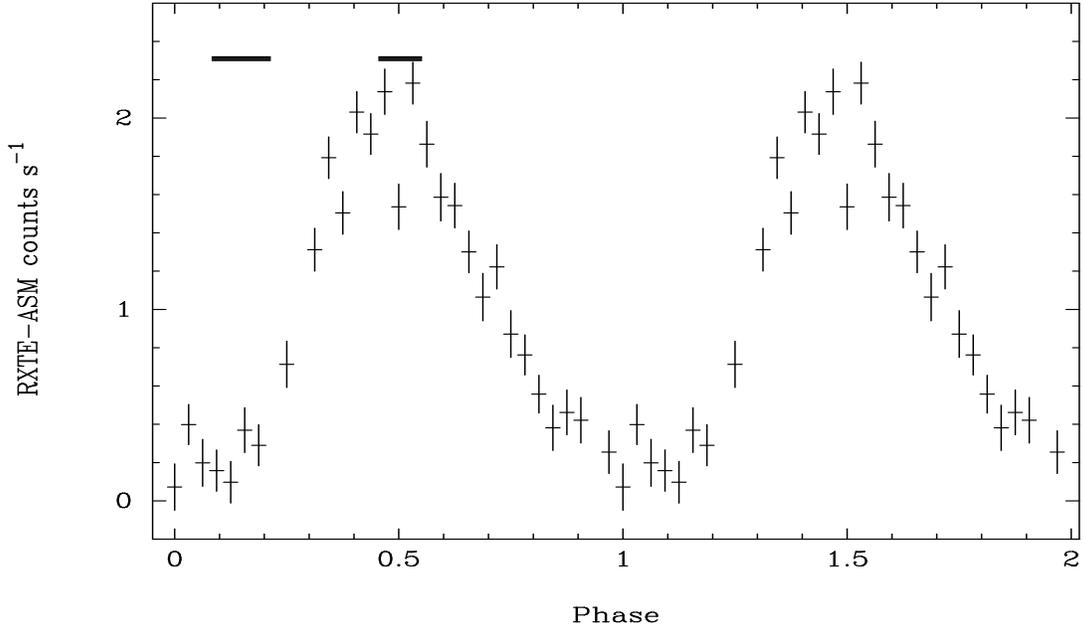}
\caption{The RXTE-ASM light curve of LMC~X-4, in 2--12 keV energy range,
folded at the long term period of 30.276 $\pm$ 0.009 days. The durations of 
the \sax\ observations with respect to the phase of the long period are 
marked at the top with bold lines.}
\label{long}
\end{figure*}

The 2--12 keV X-ray light curve of LMC~X-4 taken with the {\it RXTE}-ASM
for last six years, is shown in Figure~\ref{long}, folded at the present
super-orbital period of 30.276 days (Paul \& Kitamoto 2002).
Two observations of LMC~X-4 were made with the \sax\ narrow field
instruments during 1997 March 13 10:48 to March 15 08:17 (UT) and 1998 
October 20 22:39 to October 22 08:05 (UT) in the low 
and high states of the source respectively. The durations of these two 
observations with respect to the phase of the long period are also shown 
in Figure~\ref{long}. We have used archival data from these two observations 
in the present work.

For the present study, we have used data from the Low and Medium Energy 
Concentrator Spectrometers LECS and MECS and the Phoswich Detector System 
PDS on-board \sax\ satellite. In the 1997 observation, the hard X-ray 
instrument PDS with a larger field of view contained mostly photons from a 
nearby transient source EXO~053109--6609.2, which is only 0.5 degree away 
from LMC~X-4 (Burderi et al. 1998). The broad band spectrum (0.1--100 keV) 
from the 1998 Beppo-SAX observation has been reported by La Barbera et al. 
(2001). To obtain energy resolved pulse profiles in hard X-rays, we 
have used PDS data from the 1998 observation.

The MECS consists of two grazing incidence telescopes with imaging gas 
scintillation proportional counters in their focal planes. The LECS uses 
an identical concentrator system as the MECS, but utilizes an ultra-thin 
entrance window and a drift less configuration to extend the low-energy 
response to 0.1 keV. Time resolution of the instruments during these 
observations was 15.25 $\mu$s and energy resolutions of LECS and MECS 
are 25\% at 0.6 keV and 8\% at 6 keV respectively. The PDS consists 
of a square array of four independent NaI(Tl)/CsI(Na) phoswich scintillation 
detectors. The energy resolution of PDS instrument is 15\% at 60 keV. The 
time resolution of the PDS instruments during 1998 observation was 0.25 ms.
For a detailed description of the Beppo-SAX mission, we refer to Boella et al. 
(1997).

\section{Timing Analysis}

\begin{figure}
\vskip 9.9 cm
\includegraphics{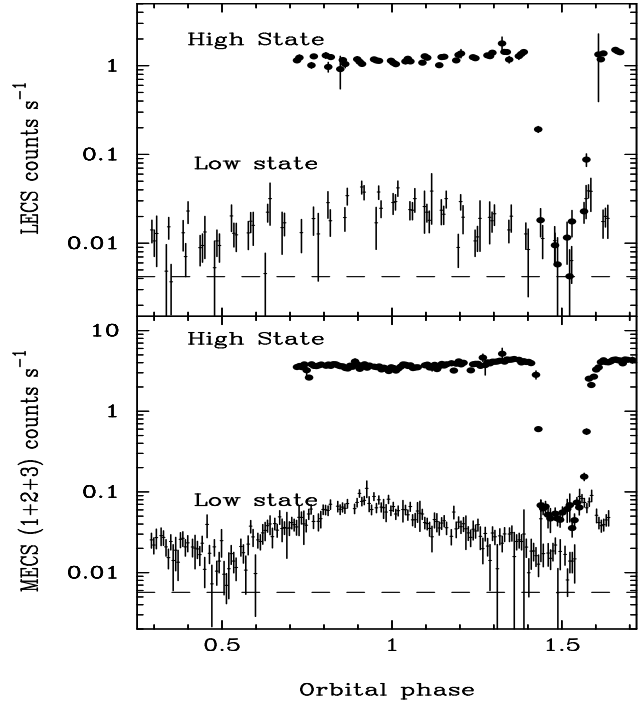}
\caption{The LECS and MECS light curves, in 0.1--10 keV and 1.3--10.0 keV 
energy ranges respectively, from the two \sax\ observations in
low and high states. The X-axis represents the orbital phase with mid-eclipse
occurring at phase 0.5 and 1.5. The background count rates, measured from
blank sky observations are shown with dashed lines. The MECS high state
observation was made with only 2 of the 3 detectors and the count rate has
been normalized to 3 detectors.}
\label{lc}
\end{figure}

The light curves of LMC~X-4 during these two \sax\ observations obtained 
with the LECS and MECS detectors are shown in Figure~\ref{lc} with respect 
to the orbital phase (mid-eclipse time = orbital phase 0.5). During the 1997
observation, which was in a low state, a smooth intensity variation is 
observed throughout the orbital period and the eclipse is not clearly seen. 
Near the end of the low state observation, an increase in count rate is
noticed which is similar to the flares observed earlier in the low state
with {\it ROSAT} (Woo et al. 1995). Similar to the other high state
observations of LMC~X-4, the 1998 observation clearly shows an eclipse
during which the X-ray intensity decreases sharply to a very low level.

\begin{figure}
\vskip 7.9 cm
\includegraphics{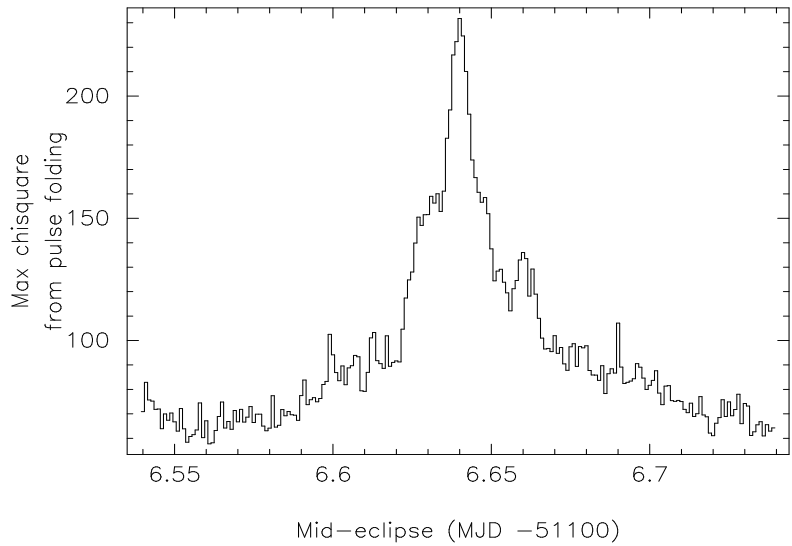}
\caption{The maximum $\chi^2$ obtained from pulse folding technique is
plotted here against the trial mid-eclipse epochs for the 1998 \sax\
observation of LMC~X-4.}
\label{res}
\end{figure}

\begin{figure}
\vskip 6.0 cm
\includegraphics{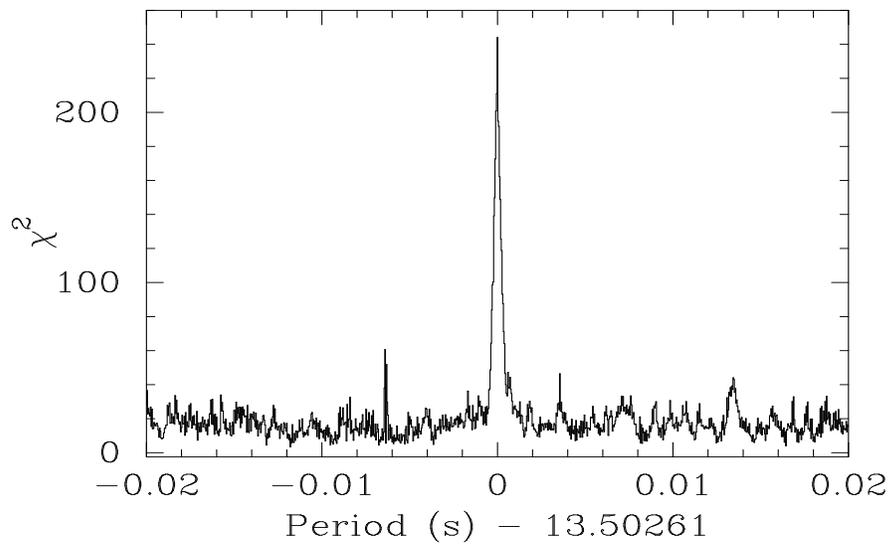}
\caption{Result of the epoch folding test on the high state MECS 
light curve of LMC~X-4.}\label{mecspp}
\end{figure}

\begin{figure}
\vskip 8.7 cm
\includegraphics{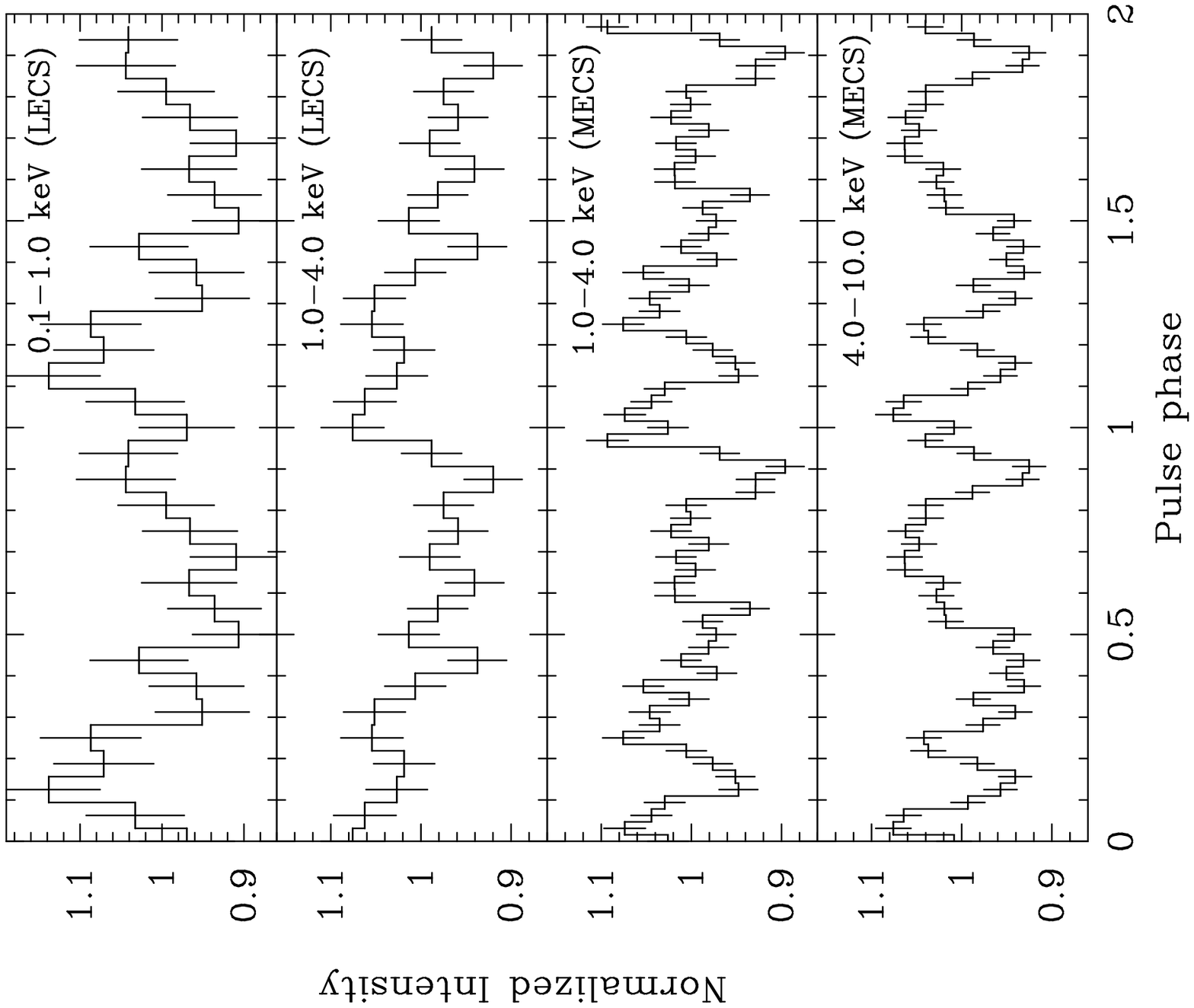}
\includegraphics{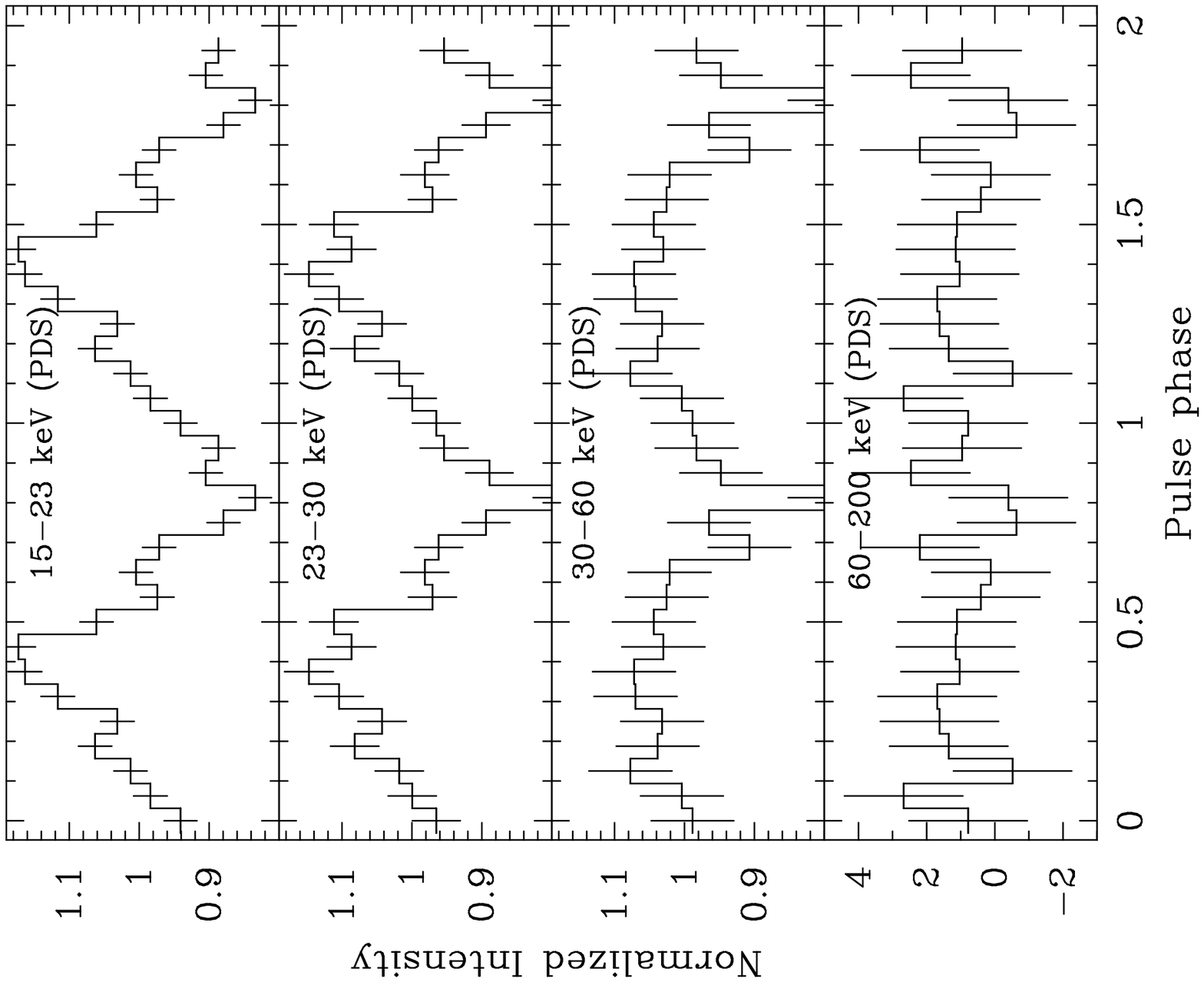}
\caption{LECS and MECS pulse profiles of LMC~X-4 in the high state are
shown here in the left panels for different energy bands with 16 and 32
phase bins per pulse respectively. Energy resolved pulse profiles from
the PDS are shown in the right panels. Two pulses are shown for clarity.}
\label{pp}
\end{figure}

For timing analysis, the arrival times of the photons were first
converted to the same at the solar system barycenter. Light curves with
time resolution of 0.25 s were extracted from circular regions of
radius $4 \arcmin$ around the source. The pulses from LMC~X-4 loose
coherence within a relatively short time scale of few thousand seconds
due to a short orbital period of 1.4 days while the semi-amplitude of
the arrival time delay due to orbital motion is 26.3 s. Additionally,
the relatively small effective area of the Beppo-SAX telescopes and 
small pulse fraction of
LMC~X-4 makes it difficult to detect pulse arrival times from small
segments of the light curve. Therefore, to detect the pulsations
and also the mid-eclipse times during these observations, we have
first corrected the light curve for the binary motion. The orbital
parameters, namely the semi-amplitude was taken to be 26.31 s and the
mid-eclipse time was derived from the quadratic solution given by Levine
et al. (2000). The binary arrival time delay correction was applied
with a range of trial mid-eclipse times around the value extrapolated
from Levine et al. (2000). Pulse folding and $\chi^2$ maximization
method was applied to all the corrected light curves. The distribution
of maximum $\chi^2$ against the trial mid-eclipse times obtained from
each pulse folding analysis is shown in Figure~\ref{res} for
the high state observation in 1998. The maximum $\chi^2$ distribution
has a Gaussian profile around the expected value of the mid-eclipse time, the 
center of which gives the correct value. Using the same method we did not
detect any pulsations in the low state. The result of the pulse folding
analysis corresponding
to the peak of the curve in Figure~\ref{res} is shown in Figure~\ref{mecspp},
which clearly shows the detection of pulsations in this observation
contrary to La Barbera et al. (2001). In the high state, the pulsations
were in fact detected independently in light curves of the LECS, MECS 
and PDS detectors. The pulse profiles obtained from the high
state light curves of the LECS, MECS, and PDS in different energy bands
are shown in Figure~\ref{pp}. From this analysis, we have derived the pulse
period of LMC~X-4 to be 13.50260(12) s and the mid-eclipse time
MJD 51106.6399(25) corresponding to orbit number 5877.

\begin{figure}
\vskip 7 cm
\includegraphics{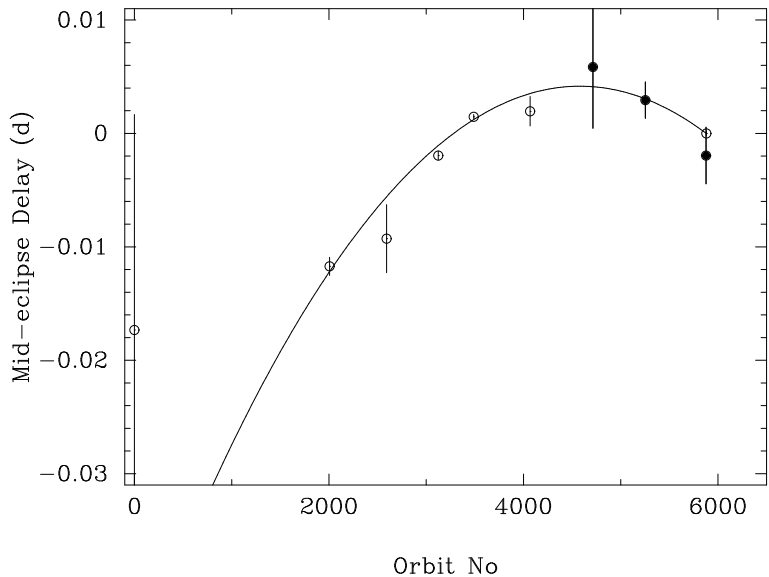}
\caption{The residual orbital epochs of LMC~X-4 relative to a constant
orbital period P = 1.40839374 d. The new measurements, after Levine et al. 2000,
are shown as filled circles. The best-fit quadratic function to the residuals is
shown as a curve.}
\label{history}
\end{figure}

Combining the new determination of mid-eclipse time with the previous
measurements (Levine et al. 2000 and references therein, Paul et al. 2002)
we fitted a second-order polynomial to derive the orbital decay rate.
Residuals of the orbital mid-eclipse time history of LMC~X-4 relative to
a constant period are shown in Figure~\ref{history}. The solid line in
the figure shows the quadratic nature of the mid-eclipse time history.
We derive a period decay rate of ${\dot{\rm P}_{\rm orb}\over {\rm P}_{\rm orb}}
= (9.89 \pm 0.05) \times 10^{-7}$ yr$^{-1}$.

The pulse profiles in different energy bands are shown in Figure~\ref{pp}.
In the low energy band (0.1--1.0 keV of LECS, top panel), it is nearly 
sinusoidal and has a complex structure in the energy band of 4.0-10.0 keV 
(MECS, bottom panel) with multiple dips superposed on a smooth sinusoidal 
profile. A phase difference between the sinusoidal pulse components of the 
two energy bands is also visible in the figure. At intermediate energies 
(1.0--4.0 keV, both LECS and MECS), the pulse profile is a mixture of the 
above two. The pulse fraction is less than 10\% and these features are 
similar to the known pulsation properties of LMC~X-4. Though the 
pulsations are seen in the PDS light curves in 15.0--60 keV energy band 
(top-right three panels of Figure~\ref{pp}), the pulse profiles do not show
the complex dipping feature, as seen in the MECS profiles, and are similar to
those obtained from the LECS light curves with certain phase difference.
The light curve above 60 keV is mainly background dominated
and we did not detect pulsations in the 60-200 keV range. 
The transient X-ray pulsar EXO 0531-66 which has a pulse
period of ~13.7 s, close to that of LMC X-4 was in the
field of view of the PDS instrument. This source was detected
in the LECS and MECS instruments at an intensity level of
about 0.2\% of LMC X-4. From a period search of the PDS
light curve in a range covering the pulse period of both
LMC X-4 and EXO 0531-66, we have verified that contamination
of the PDS folded light curve of LCM X-4 was minimal.

\section{Pulse phase averaged spectroscopy}

For spectral analysis, we have extracted LECS spectra from regions of radius 
$6 \arcmin$ centered on the object (the object was at the center of the 
field of view of both the instruments). The combined MECS source counts
(MECS~1+2+3 in the 1997 observation and MECS~2+3 in the 1998 observation) 
were extracted from circular regions with a $4 \arcmin$ radius. For spectral 
fitting, the September 1997 LECS and MECS1 response matrices were used.
Background spectra for both LECS and MECS instruments were extracted from
appropriate source-free regions of the field of view with extraction region
on the detector similar to the source extraction regions. Some rebinning 
was done to allow the use of $\chi^2$-statistic. Events were selected in 
the energy ranges 0.1--4.0 keV for LECS and 1.65--10.0 keV for MECS where 
the instrument responses are well determined. Combined spectra from the 
LECS and MECS detectors, after appropriate background subtraction, were 
fitted simultaneously. All the spectral parameters, other than the relative 
normalization, were tied to be the same for both detectors and the
minimum value of hydrogen column density $N_H$ was set at the value of 
Galactic column density in the source direction.

\subsection{Low state}

\begin{figure}
\vskip 7.5 cm
\includegraphics{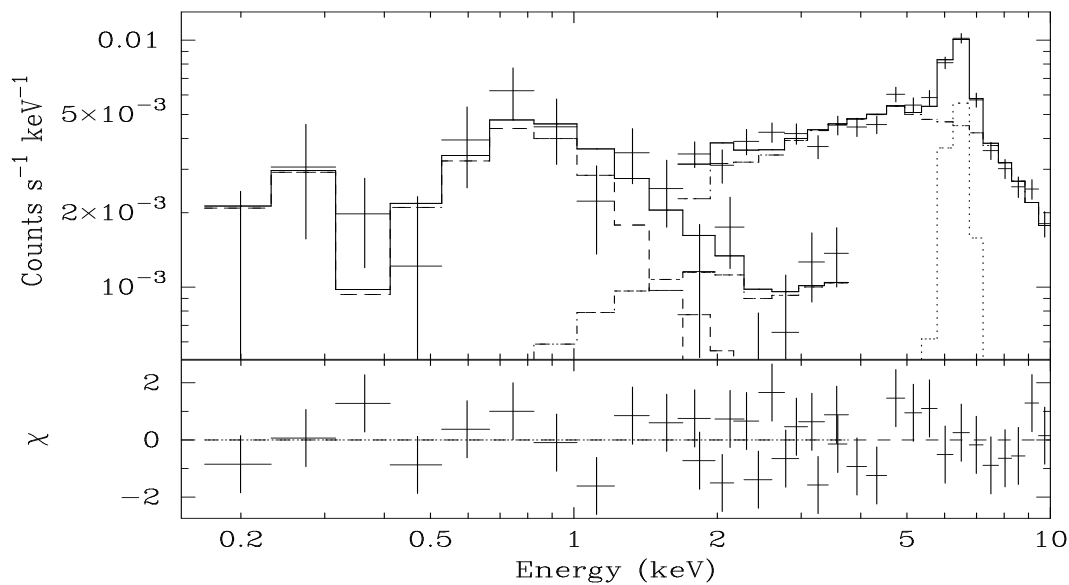}
\caption{The low state LECS and MECS count rate energy spectra in 0.1--10 keV
energy band with the folded model comprising of a thermal-bremsstrahlung, a 
hard power-law component and a broad iron emission line are shown here. The 
bottom panel shows the contributions of the residuals to the $\chi^2$ for 
each energy bin.}
\label{lss}
\end{figure}

In the low state observation of 1997, the source count rate is about two
orders of magnitude lower than the high state. Therefore, spectral fitting
required significant rebinning. The spectrum, when fitted to a single
power-law model with line of sight absorption, showed significant
soft excess below 1 keV. Addition of a soft component in the model improves
the spectral fit; we have tried two different components such as black-body
emission and bremsstrahlung component to fit the soft excess.
Results of the spectral fits are given in Table~\ref{table_low}.
From the spectral fitting, we are unable to distinguish between these two
models of the soft component. The low state photon spectrum along with
a spectral model comprising of three components, a hard power-law, a Gaussian
emission line and a bremsstrahlung emission are shown in Figure \ref{lss}. 
Two notable features of the low state spectrum are that the power-law component 
is very hard with a photon index of $\sim$ 0.1, and the iron line equivalent 
width is large, $\sim$ 1.3 keV. Both of these have been observed earlier with 
the RXTE (\cite{np:03}). The line of sight absorption during low intensity 
state is found to be $\sim$ 5 $\times$ 10$^{20}$ atoms cm$^{-2}$, similar to 
the Galactic column density, and the characteristic temperature of the soft
component is lower than that in the high state.

\subsection{High state}

\begin{figure}
\vskip 7.9 cm
\includegraphics{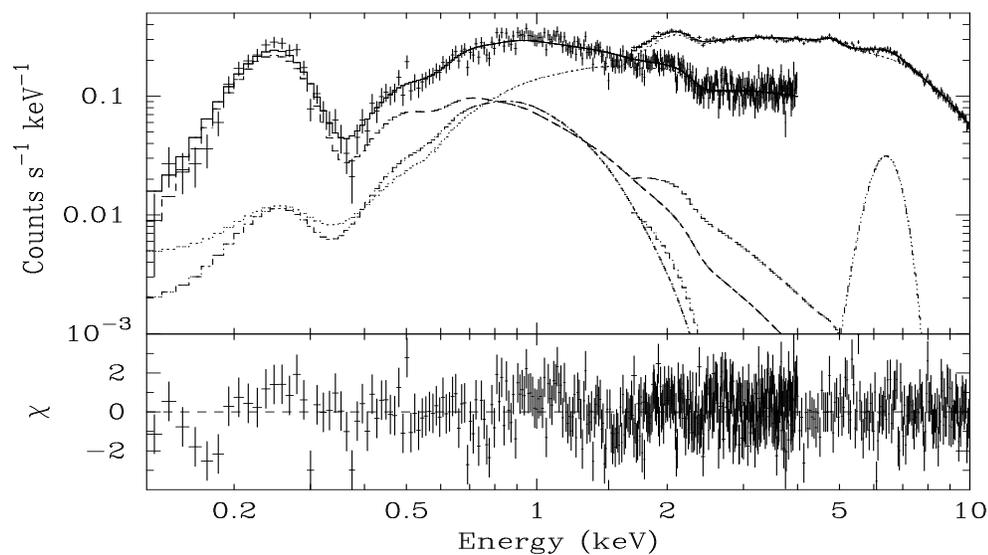}
\caption{High intensity state energy spectra of LMC~X-4 obtained with the 
LECS and MECS detectors, along with the best-fit model comprising
a hard power law component, a broad iron emission line, a soft blackbody
emission, and a soft power law. The bottom panel shows the contributions
of the residuals to the $\chi^2$ for 
each energy bin.}
\label{hss}
\end{figure}

A power-law fit to the high state spectrum shows a large soft X-ray excess.
Similar to the low state, we tried to fit the soft excess with different
single components~: black-body, thermal bremsstrahlung and soft power-law, 
none gives a satisfactory fit. Finally we found that a combination of any 
two of these components for the soft excess improves the spectral fit 
to some extent. The best-fit model is the one comprising a hard power-law 
of photon index 0.65, an iron K$_\alpha$ emission line of equivalent width 
240 eV, a black-body component of temperature 0.15 keV and a soft power-law 
with photon index $\sim$ 3. The LECS and MECS count rate spectra are shown in
Figure~\ref{hss} along with contributions of individual components.
Irrespective of the model chosen to fit the soft excess, the soft
component starts dominating the spectrum at energies below 0.8 keV.

\begin{deluxetable}{lllll}
\footnotesize
\tablecaption{Spectral parameters for LMC~X-4 during low intensity state}
\tablewidth{0pt}
\tablehead{
\colhead{Parameter} &\colhead{Model-I} &\colhead{Model-II} &\colhead{Model-III} }
\startdata
N$_H$$^1$          &5.5          &5.5$^{+4.9}$     &5.5$^{+3.0}$   \\
$\Gamma$       &0.11         &0.06$^{+0.07}_{-0.1}$   &0.11$\pm$0.06  \\
$kT$ (keV)       &------       &-----    &0.14$\pm$0.02       \\
$kT_{Br}$ (keV)	 &------       &0.4$^{+0.11}_{-0.09}$     &----- \\
Fe$_E$ (keV) $^2$ &6.45    &6.46$\pm$0.05    &6.45$^{+0.07}_{-0.04}$ \\
Fe$_W$ (keV) $^3$ &0.26    &0.26$\pm$0.08  &0.26$\pm$0.07  \\
W$_0$ (keV) $^4$  &1.26    &1.26   &1.26    \\
Model flux$^5$        &2.8     &3.1   &3.3     \\
Fe line flux $^5$     &0.39    &0.37  &0.39   \\
Reduced $\chi^2$      &2.8 (29) &1.17 (27)  &1.36 (27)  \\
\enddata
\tablenotetext{}{$^1$ : 10$^{20}$ atoms cm$^{-2}$, $^2$ : Iron line energy, $^3$ : Iron line width, $^4$ : Iron equivalent width, $^5$ : 10$^{-12}$ ergs cm$^{-2}$ s$^{-1}$}
\tablenotetext{}{Model-I = W$_{abs}$ * (Po + Gau), Model-II = W$_{abs}$ * (Br + Po + Gau), Model-III = W$_{abs}$ * (BB + Po + Gau)}
\tablenotetext{}{W$_{abs}$ = Photoelectric absorption parameterized as equivalent hydrogen column density N$_H$, Po = Power-law with photon index $\Gamma$, BB = blackbody component with temperature $kT$, Br = thermal-bremsstrahlung-type component with plasma temperature $kT_{Br}$ and Gau = Gaussian function for iron K$_\alpha$ line.}
\label{table_low}
\end{deluxetable}

\section{Pulse phase resolved spectroscopy}

Since we have not detected any pulsations in the low state observation of
1997, the pulse phase resolved spectroscopy is done only on the 1998
data set to understand the nature of the soft component. The photon arrival
times in the LECS and MECS event files were corrected for the solar system
barycenter and for the arrival time delays due to orbital motion.
Following this, spectra were accumulated into 16 pulse phases by applying
phase filtering in the FTOOLS task XSELECT. As in the case of phase-averaged
spectroscopy, the background spectra were extracted from source free regions
in the event files and appropriate response files were used for the spectral
fitting.

\begin{figure}
\vskip 7.5 cm
\includegraphics{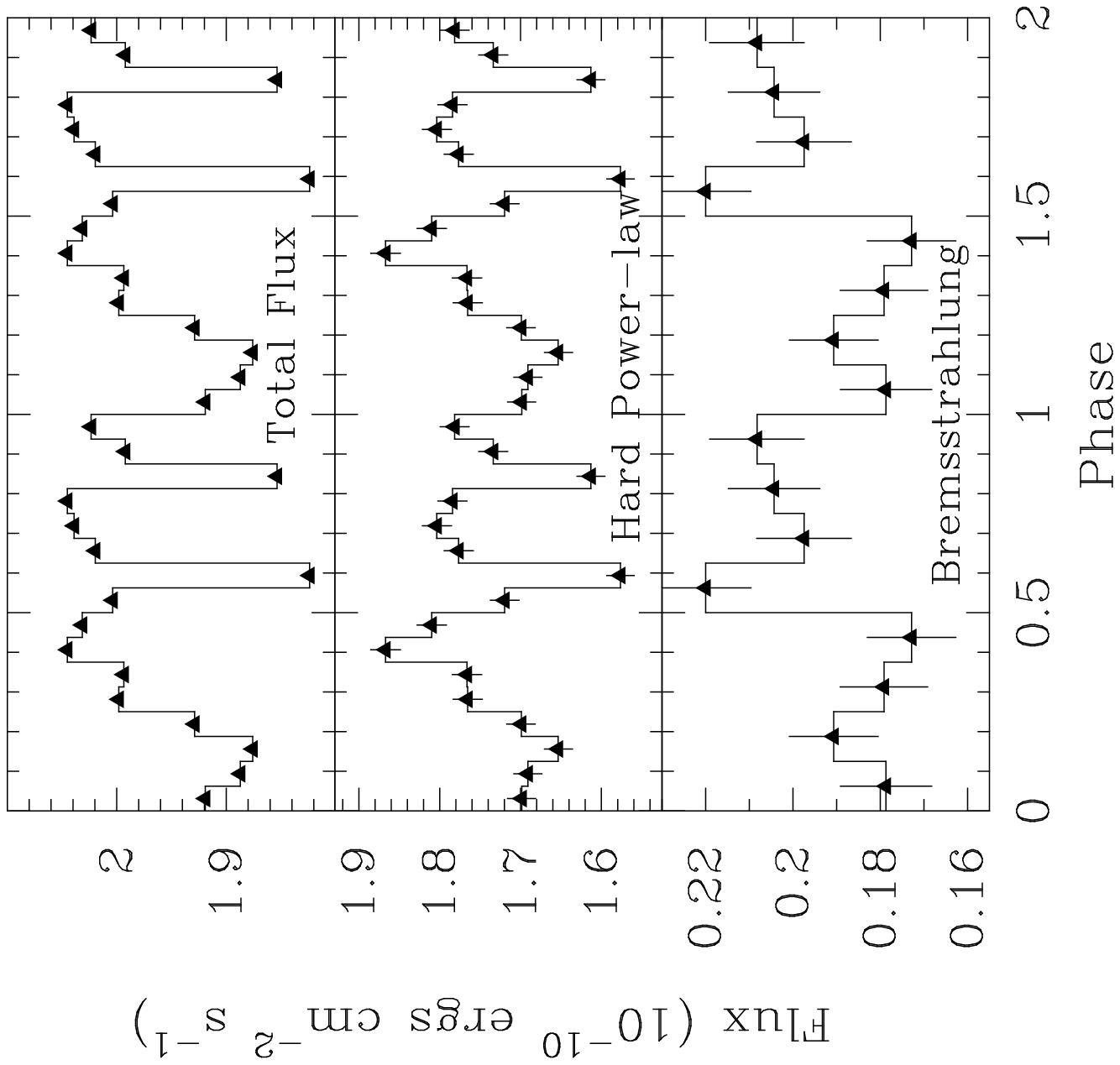}
\caption{Modulation of the total flux, hard power-law flux and flux of the 
soft spectral component in 0.1 -- 10.0 keV energy band of LMC~X-4 obtained from
the pulse-phase-resolved spectroscopy of the LECS and MECS spectra
during the high intensity state of the super-orbital period. The lower panel 
is rebinned to 8 bins per pulse for clear visibility of the pulsations in the 
soft component.}
\label{mod}
\end{figure}

Each pulse phase resolved spectra has much inferior signal to noise
ratio compared to the phase averaged spectrum. Therefore, sometimes it
is not possible to constrain all the model parameters, specially if a
complicated spectral model is used. Since our aim is to investigate
nature of the soft excess (whether pulsating or not) we used only two
models, in which a single component (either a soft power-law or a 
bremsstrahlung) is used for the soft excess and a relatively
low reduced $\chi^{2}$ is obtained. For the phase resolved spectra,
the iron-line energy, line-width and $N_H$ were fixed to their phase-averaged 
values and all the other spectral parameters were allowed to vary. The 
continuum flux and the fluxes of the soft and hard components in 0.1 -- 10.0 
keV energy range were estimated for all the 16 phase-resolved spectra. The 
modulation in the X-ray flux for the hard and soft spectral components and 
the total flux are shown in Figure~\ref{mod} along with the 1$\sigma$ error 
estimates. Pulse phase resolved spectral analysis shows that modulation of 
the hard power-law flux is very similar to the pulse profile at higher 
energies. A pulsating nature of the soft-spectral component is clearly detected 
irrespective of the spectral model used. The pulsating soft component has 
a nearly sinusoidal profile, dissimilar to the complex profile seen at higher
energies, with a certain phase difference with the hard component. These 
properties are very similar to what is seen in Her~X-1 (\cite{endo}) and 
SMC~X-1 (\cite{paul:02}).

\section{Discussion}

\subsection{Orbital evolution}

There are several ways to measure orbital ephemeris of X-ray binaries
from X-ray observations. Pulse frequency modulation or arrival time
delay measurement of X-ray pulsars gives very accurate measurement of
mid-eclipse time (T$_{\pi\over 2}$) of binary X-ray pulsars and has so
far been used to measure the orbital parameters of a large number of
binary X-ray pulsars. A quadratic nature of the mid-eclipse time history
during the last two/three decades unambiguously established an orbital
evolution with a time scale of about $10^{5}-10^{6}$ yr in several binary
X-ray pulsars (Cen~X-3, Nagase et al.  1992; LMC~X-4, Levine et al. 2000;
SMC~X-1, Wojdowski et al. 1998 etc.).  In case of low mass X-ray binary
pulsars, the orbital evolution is mainly due to
conservative mass transfer from the companion to the neutron star whereas 
in case of high mass X-ray binaries, it is either due to mass loss from
the system and/or strong tidal interaction between the two stars. The 
measurement of one new mid-eclipse times of LMC~X-4 with \sax\, and two
with {\it ASCA} (\cite{paul:02}) allows us to determine the orbital period
derivative of the source with better accuracy. It has been pointed out
before that tidal interaction is probably the dominating effect in orbital
evolution of LMC~X-4 ( Levine et al. 2000).\\

\subsection{The iron emission line}

Spectral analysis of the high and low states of LMC~X-4 showed significant
differences. Presence of a prominent iron emission line can be seen in the
low intensity state of the source (Figure~\ref{lss}). While the continuum
flux in 0.1--10 keV band in the two states differ by a factor of $\sim$ 60,
the iron line flux differs only by a factor of $\sim$ 12. This results in a
much larger equivalent width of 1.26 keV in the low state compared to 200 eV
in the high state. In view of an obscuring precessing accretion disk model
for the super-orbital period, it is interesting to note that the column
density is very similar in the two states and are close to the Galactic
value. It was suggested (\cite{hein:99}) that if the low state X-rays are
due to Compton scattering by circumstellar material, significant obscuration
by the accretion disk in the low state may still be true. Partial coverage 
of the hard X-ray emission could be a possible explanation for the low 
state. However, we found that addition of a partial covering model component 
to the hard power-law does not fit the low state spectrum.

The spectral results obtained from present work are consistent with
those obtained from the {\it RXTE}-PCA observations (\cite{np:03}). However,
a much better sensitivity for weak sources and better energy resolution of
\sax\ LECS and MECS compared to {\it RXTE}-PCA
instruments gives improved confidence in the equivalent width measurements
of the iron emission line. A smaller iron line flux in the low state
indicates that most of the line emission is produced in a region comparable
to or smaller than the size of the obscuring material, probably the
accretion disk. But an increased equivalent width of the iron line in
low state also indicates that a part of the emission line must originate
in a region further away from the neutron star. It is known that the
HMXB pulsars often show several iron line components from different
ionization species, probably produced in different regions
of the system. The best example is the case of Cen X-3 where at least
three components with different variability characteristics were resolved
(Ebisawa et al. 1996). The flux evolution of different line components
during eclipse egress suggests that the 6.4 keV line is emitted close to
the neutron star, while the other components are probably emitted in a
more extended highly ionized plasma (Nagase et al. 1992, Ebisawa et al.
1996).  In the case of LMC X-4 also, La Barbera et al. (2001) claimed
detection of more than one component with one of them being at 6.1 keV. 
High spectral resolution observations with Chandra or future ASTRO-E mission 
will allow one to find out whether these two species of iron have different 
ionization state. High spectral resolution observations in different phases 
of the super-orbital and orbital periods of LMC~X-4, including during the 
eclipse, will finally help to settle the issue regarding the origin of the iron
emission line in LMC~X-4.

\subsection{Pulsations of the soft excess and its origin}

A soft excess above the hard power-law component is now known to be
present in several accreting pulsars and the soft component is also
known to be pulsating in some of these sources. The soft excess has
been modeled with several different types of emission, but no single
model is applicable to all sources with soft excess. A black-body type
pulsating soft excess can describe the emission from Her~X-1 well
(Endo et al. 2000), however, it runs into problem in case of the
more luminous sources like SMC~X-1 and LMC~X-4 (Paul et. al. 2002).
In some sources, a bremsstrahlung type of soft emission describes the
spectrum well. However, since the emission region has to be very large
and the cooling time scale is also very large, such an emission is not
expected to pulsate. The soft excess in LMC X-4 is more complex and
more than one component is needed to model the soft excess. However,
when a bremsstrahlung component is included, it produces most of
the soft excess and the pulsating soft excess requires this component
to pulsate. LMC~X-4 has a very small pulse fraction, and the
soft excess peaks at energies lower than the {\it ASCA} threshold.
As a result, {\it ASCA} observations of LMC~X-4 were inconclusive
regarding the pulsations of the soft excess in LMC~X-4. In the present
work with \sax, we have obtained definite detection of pulsations in
the soft excess irrespective of the spectral model used.

In the low state, the LECS and MECS light curves (Figure~\ref{lc}) show
large amplitude but gradual intensity variation over the entire binary orbit.
The sharp eclipse, that is easily detectable in the high state, is not clear
in the low state at all. This indicates that in low state, the observed X-rays
are due to reprocessing from a scattering region that has size comparable
to that of the companion star. It is possible that in different orbital
phases part of the scattering region is blocked from view by the companion star.
The amplitude of variation, i.e, ratio between the maximum and minimum in
the low state binary light curve is larger in MECS compared to LECS.
This is possible if the soft excess is produced in a larger region and
the hard power-law component originated near the neutron star is heavily
absorbed with the absorption highly varying with orbital phase.
Using a radius of 8.1 R$_\odot$ of the companion star (Woo 1993) and
a binary separation of 26.31 lt-s (Levine et al. 1991), the solid angle
subtended by the companion star at the neutron star is
calculated to be $\sim$4$\pi$/7. In the present work, the soft excess
flux in the low and high states of LMC~X-4 are measured to be about about
18\% of the hard power-law flux. Therefore, one probable site for
production of the soft excess is the surface of the companion star,
which can work as a reprocessing agent. Different visibility of the
stellar surface that is irradiated by the neutron star can cause the
observed orbital soft X-ray intensity modulation in the low state.

\section*{Acknowledgments}

We thank the referee for her/his valuable suggestions that helped us 
to improve the content of this paper and also regarding analysis of the
PDS data. We thank A. R. Rao for some valuable discussions.
The Beppo-SAX satellite is a joint Italian and Dutch program. 
We thank the staff members of Beppo-SAX Science Data Center and 
RXTE/ASM group for making the data public. 

{}

\clearpage

\begin{thebibliography}{}

\bibitem[Boella et al. 1997]{boella:97}
Boella, G., Butler, R. C., Perola, G. C., Piro, L., Scarsi, L., Bleeker, 
J. A. M. 1997, A\&AS, 122, 299

\bibitem[]{488}
Boroson, B., Kallman, T., McCray, R., Vrtilek, S. D., \& Raymond, J.
1999, ApJ, 519, 191

\bibitem[Burderi et al. (1998)]{burd:98}
Burderi, L., di Salvo, T., Robba, N. R., del Sordo, S., Santangelo, A.,
\& Segreto, A. 1998, ApJ, 498, 831

\bibitem[Endo 2000]{endo:00}
Endo, T. 2000, Ph. D. thesis

\bibitem[Endo et al. 2000]{endo}
Endo, T., Nagase, F., \& Mihara, T. 2000, PASJ, 52, 223

\bibitem[Heindl et al. 1999]{hein:99}
Heindl, W. A., Gruber, D. E., Rothschild, R. E., Crannell, C. J., Lang, F. L.,
Kaplan, L. 1999, American Astronomical Society, HEAD meeting 31, 15.21

\bibitem[Ilovaisky et al. 1984]{ilovaisky84}
Ilovaisky, S. A., Chevalier, C., Motch, C., Pakull, M., van Paradijs, J.,
\& Lub, J. 1984, A\&A, 140, 251

\bibitem[]{510}
Katz, J. I. 1973, NPhS, 246, 87

\bibitem[Kelley et al. 1983]{kelley:83}
Kelley, R. L., Jernigan, J. G., Levine, A., Petro, L. D., \& Rappaport, S.
1983, ApJ, 264, 568

\bibitem[Kohno et al. 2000]{kohno:00}
Kohno, M., Yokogawa, Jun, \& Koyama, K 2000, PASJ, 52, 299

\bibitem[Lang et al. 1981]{lang:81}
Lang, F. L. et al. 1981, ApJ, 246, L21

\bibitem[Leahy 2001]{leahy:01}
Leahy, D. A. 2001, ApJ, 547, 449

\bibitem[Levine et al. 1991]{levine:91}
Levine, A., Rappaport, S., Putney, A., Corbet, R., \& Nagase, F.
1991, ApJ, 381, 101

\bibitem[Levine et al. 2000]{levi:00}
Levine, A. M., Rappaport, S. A., \& Zojcheski, G. 2000, ApJ, 541, 194

\bibitem[Li, Rappaport \& Epstein 1978]{li:78}
Li, F., Rappaport, S., \& Epstein, A. 1978, Nature, 271, 37

\bibitem[Nagase et al. 1992]{nagase:92}
Nagase, F., Corbet, R. H. D., Day, C. S. R., Inoue, H., Takeshima, T.,
Yoshida, K., \& Mihara, T. 1992, ApJ, 396, 147

\bibitem[Nagase et al. 1986]{nagase:86}
Nagase, F., Hayakawa, S., Sato, N., et al. 1986, PASJ, 38, 547

\bibitem[Naik \& Paul 2003]{np:03}
Naik, S., \& Paul, B. 2003, A\&A, 401, 265

\bibitem[Paul \& Kitamoto 2002]{pk:02}
Paul, B., \& Kitamoto, S. 2002, JApA, 22, 33

\bibitem[Paul et al. 2002]{paul:02}
Paul, B., Nagase, F., Endo, T., Dotani, T., Yokogawa, J., \& Nishiuchi, M. 2002, ApJ, 579, 411

\bibitem[Pietsch et al. 1985]{pietsch:85}
Pietsch, W., Voges, W., Pakull, M., \& Staubert, R. 1985, SSRv, 40, 371

\bibitem[]{555}
Roberts, M. S. 1974, Sci, 183, 371

\bibitem[]{561}
Vrtilek, S. D., Boroson, B., Cheng, F. H., McCray, R., \& Nagase, F.
1997, ApJ, 490, 377

\bibitem[Yokogawa et al. 2000]{yokogawa:00}
Yokogawa, J., Paul, B., Ozaki, M., Nagase, F., Chakrabarty, D., Takeshima, T. 2000, ApJ, 539, 191

\bibitem[White 1978]{whit:78}
White, N. E. 1978, Natur, 271, 38

\bibitem[Wojdowski et a. 1998]{wojd:98}
Wojdowski, P., Clark, G. W., Levine, A. M., Woo, J. W., \& Zhang, S. N.,
1998, ApJ, 502, 253

\bibitem[Woo 1993]{woo:93}
Woo, J. W. 1993, PhDT, 26

\bibitem[Woo et al. 1995]{woo:95}
Woo, J. W., Clark, G. W., \& Levine, A. M. 1995, ApJ, 449, 880

\bibitem[Woo et al. 1996]{woo:96}
Woo, J. W., Clark, G. W., Levine, A. M., Corbet, R. H. D., \&
Nagase, F. 1996, ApJ, 467, 811
\end{thebibliography}
\end{document}